\documentclass[preprint]{emulateapj}
\usepackage{color}
\usepackage{amsmath,natbib,graphicx}
\usepackage{epsf}
\usepackage{epstopdf}
\input{epsf}
\usepackage{epsfig}
\usepackage{color}
\DeclareGraphicsExtensions{.jpg,.pdf,.png,.eps,.ps}

\def\red#1 {\textcolor{red}{#1}\ }

\def\gs{\mathrel{\raise0.35ex\hbox{$\scriptstyle >$}\kern-0.6em \lower0.40ex\hbox{{$\scriptstyle \sim$}}}}
\def\ls{\mathrel{\raise0.35ex\hbox{$\scriptstyle <$}\kern-0.6em \lower0.40ex\hbox{{$\scriptstyle \sim$}}}}
\newcommand{\Msolar}{\mbox{$M_{\odot}\,$}}
\newcommand{\Lsolar}{\mbox{$L_{\odot}\,$}}
\newcommand{\arcmins}{\mbox{$^{\prime}$}}
\newcommand{\arcsecs}{\mbox{$^{\prime\prime}$}}

\shorttitle{Submillimeter Observations of SPT Sources}
\shortauthors{Greve et al.}

\begin{document}


\title{Submillimeter observations of millimeter bright galaxies discovered by the South Pole Telescope}


\author{
 T.~R. Greve,\altaffilmark{1}	
 J.~D. Vieira,\altaffilmark{2}
 A.~Wei\ss,\altaffilmark{3}
 J.~E.~Aguirre,\altaffilmark{4}
 K.~A.~Aird,\altaffilmark{5}
 M.~L.~N.~Ashby,\altaffilmark{6}
 B.~A.~Benson,\altaffilmark{7,8}
 L.~E.~Bleem,\altaffilmark{7,9}
 C.~M.~Bradford,\altaffilmark{10}
 M.~Brodwin,\altaffilmark{6}
 J.~E.~Carlstrom,\altaffilmark{7,8,9,11,12}
 C.~L.~Chang,\altaffilmark{7,8,12}
 S.~C.~Chapman,\altaffilmark{13}
 T.~M.~Crawford,\altaffilmark{7,11}
 C.~de~Breuck\altaffilmark{14}
 T.~de~Haan,\altaffilmark{15}
 M.~A.~Dobbs,\altaffilmark{15}
 T.~Downes,\altaffilmark{2}
 C.~D.~Fassnacht,\altaffilmark{16}
 G.~Fazio,\altaffilmark{6}
 E.~M.~George,\altaffilmark{17}
 M.~Gladders,\altaffilmark{7,11}
 A.~H.~Gonzalez,\altaffilmark{18} 
 N.~W.~Halverson,\altaffilmark{19} 
 Y.~Hezaveh,\altaffilmark{15}
 F.~W.~High,\altaffilmark{7,11}
 G.~P.~Holder,\altaffilmark{15}
 W.~L.~Holzapfel,\altaffilmark{17}
 S.~Hoover,\altaffilmark{7,8}
 J.~D.~Hrubes,\altaffilmark{5}
 M.~Johnson,\altaffilmark{20}
 R.~Keisler,\altaffilmark{7,9}
 L.~Knox,\altaffilmark{16}
 A.~T.~Lee,\altaffilmark{17,21}
 E.~M.~Leitch,\altaffilmark{7,11}
 M.~Lueker,\altaffilmark{2}
 D.~Luong-Van,\altaffilmark{5}
 M.~Malkan,\altaffilmark{20}
 D.~P.~Marrone,\altaffilmark{22}
 V.~McIntyre,\altaffilmark{23} 
 J.~J.~McMahon,\altaffilmark{24} 
 J.~Mehl,\altaffilmark{7}
 K.~M.~Menten,\altaffilmark{3}
 S.~S.~Meyer,\altaffilmark{7,8,9,11}
 T.~Montroy,\altaffilmark{25} 
 E.~J.~Murphy,\altaffilmark{26} 
 T.~Natoli,\altaffilmark{7,8}
 S.~Padin,\altaffilmark{2,7,11}
 T.~Plagge,\altaffilmark{7,11}
 C.~Pryke,\altaffilmark{7,8,11}
 C.~L.~Reichardt,\altaffilmark{17}
 A.~Rest,\altaffilmark{27,28} 
 M.~Rosenman,\altaffilmark{4}
 J.~Ruel,\altaffilmark{27} 
 J.~E.~Ruhl,\altaffilmark{25} 
 K.~K.~Schaffer,\altaffilmark{7,29} 
 K.~Sharon,\altaffilmark{7}
 L.~Shaw,\altaffilmark{30} 
 E.~Shirokoff,\altaffilmark{2,17}
 B.~Stalder,\altaffilmark{6,27} 
 S.~A.~Stanford,\altaffilmark{16}
 Z.~Staniszewski,\altaffilmark{2,25} 
 A.~A.~Stark,\altaffilmark{6}
 K.~Story,\altaffilmark{7,8}
 K.~Vanderlinde,\altaffilmark{15}
 W.~Walsh,\altaffilmark{6}
 N.~Welikala,\altaffilmark{31} 
 R.~Williamson\altaffilmark{7,11}
}

\altaffiltext{1}{
Department of Physics and Astronomy, University College London, Gower Street, London WC1E 6BT, UK}
\altaffiltext{2}{California Institute of Technology, 1216 E. California Blvd., Pasadena, CA 91125, USA}
\altaffiltext{3}{Max-Planck-Institut f\"{u}r Radioastronomie,
Auf dem H\"{u}gel 69 D-53121 Bonn, Germany}
\altaffiltext{4}{University of Pennsylvania,
209 South 33rd Street 
Philadelphia, PA 19104, USA}
\altaffiltext{5}{University of Chicago,
5640 South Ellis Avenue, Chicago, IL 60637, USA}
\altaffiltext{6}{Harvard-Smithsonian Center for Astrophysics,
60 Garden Street, Cambridge, MA 02138, USA}
\altaffiltext{7}{Kavli Institute for Cosmological Physics,
University of Chicago, 5640 South Ellis Avenue, Chicago, IL 60637, USA}
\altaffiltext{8}{Enrico Fermi Institute,
University of Chicago,
5640 South Ellis Avenue, Chicago, IL 60637, USA}
\altaffiltext{9}{Department of Physics,
University of Chicago,
5640 South Ellis Avenue, Chicago, IL 60637, USA}
\altaffiltext{10}{Jet Propulsion Laboratory, 4800 Oak Grove Drive, Pasadena, CA 91109, USA}
\altaffiltext{11}{Department of Astronomy and Astrophysics,
University of Chicago,
5640 South Ellis Avenue, Chicago, IL 60637, USA}
\altaffiltext{12}{Argonne National Laboratory, 9700 S. Cass Avenue, Argonne, IL 60439, USA}
\altaffiltext{13}{Institute of Astronomy, University of Cambridge, Madingley Road, Cambridge CB3 0HA, UK}
\altaffiltext{14}{European Southern Observatory, Karl-Schwarzschild Strasse, D-85748 Garching bei M\"unchen, Germany}
\altaffiltext{15}{Department of Physics,
McGill University, 3600 Rue University, 
Montreal, Quebec H3A 2T8, Canada}
\altaffiltext{16}{Department of Physics, 
University of California, One Shields Avenue, Davis, CA 95616, USA}
\altaffiltext{17}{Department of Physics,
University of California, Berkeley, CA, 94720, USA}
\altaffiltext{18}{Department of Astronomy, University of Florida,
Gainesville, FL 32611, USA}
\altaffiltext{19}{Department of Astrophysical and Planetary Sciences and Department of Physics,
University of Colorado,
Boulder, CO 80309, USA}
\altaffiltext{20}{Department of Physics and Astronomy, University of California, Los Angeles, CA 90095-1547, USA}
\altaffiltext{21}{Physics Division,
Lawrence Berkeley National Laboratory,
Berkeley, CA, 94720, USA}
\altaffiltext{22}{Steward Observatory, University of Arizona, 933 North Cherry Avenue, Tucson, AZ 85721, USA}
\altaffiltext{23}{Australia Telescope National Facility, CSIRO, Epping, NSW 1710, Australia}
\altaffiltext{24}{Department of Physics, University of Michigan, 450 Church Street, Ann  
Arbor, MI 48109, USA}
\altaffiltext{25}{Physics Department, Center for Education and Research in Cosmology 
and Astrophysics, Case Western Reserve University, Cleveland, OH 44106, USA}
\altaffiltext{26}{Observatories of the Carnegie Institution for Science, 813 Santa Barbara Street, Pasadena, CA 91101, USA}
\altaffiltext{27}{Department of Physics, Harvard University, 17 Oxford Street, Cambridge, MA 02138, USA}
\altaffiltext{28}{Space Telescope Science Institute, 3700 San Martin Dr., Baltimore, MD 21218, USA}
\altaffiltext{29}{Liberal Arts Department, School of the Art Institute of Chicago, 112 S Michigan Ave, Chicago, IL, USA 60603}
\altaffiltext{30}{Department of Physics, Yale University, P.O. Box 208210, New Haven,
CT, USA 06520-8120}
\altaffiltext{31}{Insitut d'Astrophysique Spatiale, B\^atiment 121, Universit\'e Paris-Sud XI \& CNRS, 91405 Orsay Cedex, France}

\email{tgreve@star.ucl.ac.uk}

\begin{abstract}
We present APEX SABOCA $350\,\rm{\mu m}$ and LABOCA $870\,\rm{\mu m}$
observations of 11 representative examples of the rare, extremely bright
($S_{\rm 1.4 mm}>15\,\rm{mJy}$), dust-dominated millimeter-selected galaxies
recently discovered by the South Pole Telescope (SPT). All 11 sources are
robustly detected with LABOCA with $40<S_{\rm 870\mu m}<130\,\rm{mJy}$,
approximately an order of magnitude higher than the canonical submillimeter galaxy (SMG)
population. Six of the sources are also detected by SABOCA at $> 3\sigma$, with the
detections or upper limits providing a key constraint on the shape of the
spectral energy distribution (SED) near its peak. We model the SEDs of these
galaxies using a simple modified blackbody and perform the same analysis on
samples of SMGs of known redshift from the literature. These calibration
samples inform the distribution of dust temperature for similar SMG
populations, and this dust temperature prior allows us to derive photometric
redshift estimates and far infrared luminosities for the sources. We find a
median redshift of $\overline{z} = 3.0$, higher than the $\overline{z}=2.2$
inferred for the normal SMG population. We also derive the apparent size of
the sources from the temperature and apparent luminosity, finding them to
appear larger than our unlensed calibration sample, which supports the idea
that these sources are gravitationally magnified by massive structures along
the line of sight.
\end{abstract}


\keywords{galaxies: high-redshift --- galaxies: formation --- galaxies: evolution --- galaxies: starbursts --- submillimeter ---   gravitational lensing}



\section{Introduction}
The first extragalactic surveys at submillimeter (submm) wavelengths carried
out at $850\,\rm{\mu m}$ with SCUBA \citep{holland99}  over a decade ago
\citep{smail97,hughes98,barger98} discovered a population of optically faint,
submm bright galaxies. Multi-wavelength follow-up campaigns subsequently
established that the bright ($S_{\rm 850\mu m}\sim 3-15\,\rm{mJy}$)
submillimeter galaxies (SMGs) represent a significant population of distant ($z
\simeq 1-3$) dust-enshrouded, far-infrared (FIR) luminous ($L_{\rm FIR}\sim 10^{13}\,\Lsolar$)
galaxies, in which large gas reservoirs ($M({\rm H_2}) \sim 10^{11}\,\Msolar$)
are being turned into stars at a prodigious rate ($SFR \sim
1000\,\Msolar\,$yr$^{-1}$; e.g., \citealt{neri03,chapman05}). 

Ground-based extragalactic surveys conducted at wavelengths of $\rm
850-1200\,\mu m$ have identified a few hundred sources in total sky area of
$\sim 4\,{\rm deg^2}$ \citep[e.g.,][]{coppin06,perera08,austermann09,weiss09}.  The
brightest sources found in these surveys have $S_{\rm 850\mu
m}\lesssim20\,{\rm mJy}$, and a source density of $\sim1\,{\rm deg^{-2}}$.  A
handful of brighter sources have been discovered in small area surveys by
targeting massive galaxy clusters and thereby taking advantage of gravitational 
magnification. For example, ultra-bright SMGs have been found behind the Bullet
cluster (1ES~0657$-$56, $S_{\rm 870\mu m}=48\,{\rm mJy}$, $z=2.79$;
\citealt{wilson08b,gonzalez10,johansson10}) and MACS\,J2135$-$010217
(SMM\,J2135$-$0102, hereafter called the ``Eyelash", $S_{\rm 870\mu
m}=106\,{\rm mJy}$, $z=2.33$; \citealt{swinbank10,ivison10,danielson11}).
Sources lensed by intervening galaxies have also been found serendipitously
($S_{\rm 870\mu m}\sim100\,{\rm mJy}$, $z=3-4$;
\citealt{lestrade10,ikarashi11}). These objects are important as
gravitationally magnified windows on the obscured star formation process in the
early universe, but large samples of these rare objects cannot be obtained in
surveys of a few square degrees. 

The landscape of submm surveys has changed dramatically in the last few years
with the advent of observatories capable of mapping large areas of sky
simultaneously in multiple bands. The Balloon-borne Large-Aperture
Submillimeter Telescope (BLAST -- \citep{pascale08}) mapped over $10\,{\rm deg^2}$ 
at $250, 350$, and $500\,{\rm \mu m}$ \citep{devlin09}, identifying many dusty
galaxies and providing photometric redshift estimates. The $10\,{\rm m}$ South
Pole Telescope (SPT) surveyed $200\,{\rm deg^2}$ to mJy depth at 1.4 and $2.0\,{\rm
mm}$ in 2008, discovering a population of rare ($\sim 0.1\,{\rm deg^{-2}}$) and
extremely bright ($> 20\,{\rm mJy}$ at $1.4\,{\rm mm}$) dusty galaxies
\citep{vieira10}.  Early results from the Science Demonstration Phase (SDP)
observations of the {\it Herschel} Astrophysical Terahertz Large Area Survey
(H-ATLAS -- \citealt{eales10}) and the {\it Herschel} Multi-tiered
Extragalactic Survey (HerMES -- \citealt{oliver10}) -- each surveying tens of
deg$^{2}$ at $250, 350$, and $500\,{\rm \mu m}$ -- also identified analogous
populations of bright sources \citep{negrello10,conley11}.

In this paper, we denote SMGs with $S_{\rm 850\mu m}>30\,{\rm mJy}$,
primarily discovered in the large area surveys described above, as
``ultra-bright" SMGs. These sources appear more luminous than galaxies
identified in smaller surveys ($S_{\rm 850\mu m}\simeq 3-15\,{\rm mJy}$); we
refer to the fainter, un-lensed, population as ``normal" SMGs.  The apparent
luminosity of the ultra-bright SMGs, and their excess relative to the expected
high-flux number counts of normal SMGs, is a strong indication that a
significant fraction are strongly lensed with magnification factors of 
order $10$ to $50$ \citep{blain96,blain99b, negrello07, hezaveh11}. 

Measuring this lensing amplification -- using high-resolution follow-up imaging
and detailed lens modeling -- is one of the keys to determining the intrinsic
properties of these ultra-bright SMGs and their place in the overall scheme of
galaxy evolution. The other key, which we focus on in this work, is determining
their redshifts. Recent ``blind" CO line searches, i.e., without prior
optical/near-IR spectroscopy, with the Zpectrometer \citep{frayer11} on the
Green Bank Telescope and Z-Spec \citep{bradford09} on the Caltech Submillimeter
Observatory yielded redshifts in the range $z\simeq 1.6-3.0$ for five
ultra-bright H-ATLAS sources \citep{frayer11, lupu10}. At present, robust
spectroscopic redshifts have been published for ten such ultra-bright SMGs --
discovered either from the ground or with {\it Herschel} -- although several
more sources from H-ATLAS and HerMES have now been spectroscopically confirmed
\citep{harris12}. The bulk of these lie within the the observed redshift range
of normal, radio-identified SMGs \citep[$z\sim 1-3$,][]{chapman05}, which may
in part be due to the frequency coverage of Zpectrometer and Z-Spec favoring $z
< 4$ CO detections. The redshift of the single $z > 4$ ultra-bright SMG
published to date was made using the IRAM PdBI WideX correlator \citep{cox11}.
A large overlap between the SPT and \textit{Herschel} sources is expected, but
the longer SPT selection wavelength ($1.4\,{\rm mm}$) does predict a broader
redshift range than the $350-500\,{\rm \mu m}$-selected {\it Herschel} sources.
The SPT may also be sensitive to a population of cooler sources that are
invisible to the shorter-wavelength {\it Herschel} selection.  The large SPT
survey area ($2500\,{\rm deg^2}$, compared to the $600$ and $380\,{\rm deg^2}$
that H-ATLAS and HerMES cover) ensures the identification of the rarest and
most highly magnified objects in the sky.

In this paper we present the first $350$ and $870\,{\rm \mu m}$ maps of a
subset of 11 SPT sources. In \S\ref{section:observations} the SABOCA and LABOCA
observations and data reduction are outlined. In \S\ref{subsection:maps} and
\ref{subsection:fluxes} we describe the submm maps, source morphology, and
fluxes, while in \S\ref{subsection:SEDs} and \ref{subsection:mBB} we fit
spectral energy distributions (SEDs), derive photometric redshifts, FIR
luminosities, and dust temperatures based on the SEDs. Finally, in
\S\ref{section:discussion} we discuss the derived properties of SPT sources, including
their redshifts, and the implications these findings have. Throughout, we
adopt a flat cosmology with $\rm \Omega_M = 0.27, \Omega_{\Lambda} = 0.73$, and
$\rm h = 0.71$ \citep{wmap1}.

\section{Observations and data reduction}\label{section:observations}

\subsection{South Pole Telescope Selection}\label{subsection:sptobs}
The SPT \citep{carlstrom11}, a 10-meter off-axis Gregorian design with a
$1\,{\rm deg^2}$ field of view, has been surveying the mm-wave sky with
unprecedented sensitivity and angular resolution since its commissioning in
2007. The SPT is located within $1\,{\rm km}$ of the geographical South Pole.
At an altitude of $2800\,{\rm m}$ above sea level, the South Pole is one of
the premier locations for mm-wave astronomy. The high altitude and low
temperatures ensure an atmosphere with low water-vapor content and excellent
transparency. Meanwhile, the location at the Earth's rotational axis allows
24-hour access to the target fields.  

The first receiver mounted on the SPT is a camera consisting of 840
transition-edge-sensor bolometers, optimized for fine-scale anisotropy studies
of the cosmic microwave background (CMB) and the discovery of distant massive
galaxy clusters through the thermal Sunyaev-Zel'dovich (SZ) effect
\citep{sunyaev72}. The 840 bolometers are split into six wedges each
containing 140 detectors. In 2008 (the season during which the SPT data in this
work was taken), the array consisted of a single $3.2\,{\rm mm}$ wedge, three
$2.0\,{\rm mm}$ wedges, and two $1.4\,{\rm mm}$ wedges. The $3.2\,{\rm mm}$
wedge did not produce science-quality data, but the $1.4$ and $2.0\,{\rm mm}$
wedges performed to specification, resulting in r.m.s.\ survey depths of
approximately $1.3\,{\rm mJy}$ at $2.0\,{\rm mm}$ and $3.4\,{\rm mJy}$ at
$1.4\,{\rm mm}$. The 10-meter primary mirror of the SPT 
results in beam sizes (FWHM) of approximately $1.0^\prime$ at
$1.4\,{\rm mm}$ and $1.1^\prime$ at $2.0\,{\rm mm}$. The camera was upgraded
to provide sensitivity in all three bands for observations starting in 2009.
The $2500\,{\rm deg^2}$ SPT-SZ survey was completed in all three bands in November 2011.

The data reduction pipeline applied to the SPT data is described in
\citet[][hereafter V10]{vieira10}. Briefly, the SPT time-ordered data from
every working detector are calibrated and bandpass filtered, one or more common
modes are removed from the data of all the detectors on a given wedge, and the
data from all detectors in a given wavelength band are co-added into a map
using inverse-noise weighting.  

Sources are identified in SPT maps as described in V10. Objects are identified
by first convolving the single-band map with a matched filter
\citep[e.g.,][]{haehnelt96} that de-weights noise and astrophysical signals on
large scales and maximizes sensitivity to point-like objects, then searching
the filtered map for high-significance peaks. Sources are then cross matched
between the 1.4 and $2.0\,{\rm mm}$ catalogs, and their fluxes are deboosted
according to the method detailed in \citet{crawford10}. Sources detected in
SPT maps are classified as dust-dominated or synchrotron-dominated based on the
ratio of their flux in the $1.4\,{\rm mm}$ and $2.0\,{\rm mm}$ bands.
Approximating the spectral behavior of sources between $1.4$ and $2.0\,{\rm
mm}$ as a power law, $S(\lambda) \propto \lambda^{-\alpha}$, we estimate the
spectral index $\alpha$ for every source and use $\alpha=1.66$ as the dividing
line between dust- and synchrotron-dominated populations, thereby removing all
flat spectrum radio quasars from the sample (see V10 for details). We apply an
additional selection on the sample by imposing a cut on sources found in the
Infrared Astronomy Satellite Faint-Source Catalog
\citep[IRAS-FSC,][]{moshir92}. This corresponds to cutting out sources with
$S_{\rm 60\mu m}>200\,{\rm mJy}$, which should remove any source at $z<1$ from
this sample. Absolute calibration for both the $1.4$ and $2.0\,{\rm mm}$ bands
is derived from the CMB and the calibration uncertainty is $\ls 10\%$, as
described in \citet{vieira10}.

For the study presented here, 11 sources from $200\,{\rm deg^2}$ of 2008 survey data
were imaged at $870$ and $350\,{\rm \mu m}$. Their $1.4\,{\rm mm}$ fluxes
range from $17$ to $40\,{\rm mJy}$ and approximate a flux limited sample. 
The sources are listed with
their full source names in Table \ref{table:sources}; throughout the paper we
refer to them by their truncated coordinates (e.g., SPT-S\,J233227$-$5358.5
becomes SPT\,2332$-$53).

\subsection{APEX submm continuum imaging}\label{subsection:apexobs}
The submm observations presented in this paper were carried out at $\rm 350$
and $\rm 870\,\mu m$ with the Submillimeter Apex BOlometer CAmera (SABOCA) and
the Large Apex BOlometer CAmera (LABOCA) at the Atacama Pathfinder EXperiment
(APEX).\footnote{APEX is a collaboration between the Max-Planck Institute f\"ur
Radioastronomie, the European Southern Observatory, and the Onsala Space
Observatory.} The sources observed and their coordinates are given in Table
\ref{table:sources}.

LABOCA is a 295-element bolometer array (Siringo et al.\ 2009) with an $\rm
11.4\arcmins$ field-of-view and a measured angular resolution of $\rm
19.7\arcsecs$ (FWHM). The center frequency of LABOCA is $345\,{\rm GHz}$
($870\,{\rm \mu m}$) with a passband FWHM of $\sim 60\,{\rm GHz}$. The
measured noise performance for these observations was $60\,{\rm mJy\,s^{1/2}}$.   

SABOCA is a 39-element transition edge sensor (TES) bolometer array (Siringo et
al.\ 2010) with a $\rm 1.5\arcmins$ field-of-view and a measured angular
resolution of $\rm 7.8\arcsecs$ (FWHM). The center frequency of SABOCA is
$860\,{\rm GHz}$ ($350\,{\rm \mu m}$) with a passband FWHM of $\sim 120\,{\rm
GHz}$. The measured noise performance for these observations was $150\,{\rm
mJy\,s^{1/2}}$.   

SABOCA and LABOCA observations of seven of our sources were carried out during
Max-Planck observing time in May 2010 (PI: Wei\ss).  A further four sources were observed
in August 2010 under ESO programs 086.A-1002A and 087.A-0968A (PI: Greve).
This brings the total number of sources presented in this paper to 11. All
SABOCA observations were obtained in excellent weather conditions (precipitable
water vapor, $PWV$, less than  $0.7\,{\rm mm}$). The LABOCA $850\,{\rm \mu m}$
observations of all targets were done in good weather conditions ($PWV <
1.5\,{\rm mm}$).

Typical integration times for LABOCA were $1\,{\rm h}$ on-source resulting in an
r.m.s.\ level of $\sim6\,{\rm mJy\,beam^{-1}}$, adequate to detect any of the
SPT sources at $z<9$, assuming a point-like source (with respect to the LABOCA
beam) and an SED similar to other SMGs.

At the time of our observations, we had no information on source redshifts, SED
shapes, or resolved structure, all of which can dramatically vary the expected
$350\,{\rm \mu m}$ flux density for a given $1.4\,{\rm mm}$ flux density. The
SABOCA observations therefore targeted a fixed r.m.s.\ of $30\,{\rm
mJy\,beam^{-1}}$ ($\sim$3\,h on-source). At this noise level we expected a
5-$\sigma$ detection for a typical redshift of $z\sim3$, assuming an unresolved
point source. For sources that were not detected at this noise level we
increased the integration time up to 5\,h on-source, yielding a r.m.s.\ of
$\sim20\,{\rm mJy\,beam^{-1}}$, which should detect a typical source out to
approximately $z\sim4$. 

Mapping was performed using the raster--spiral mode \citep{siringo09} for both
bolometer arrays. This mode yields map sizes slightly larger than the field of
view of the arrays ($\approx 1.5\arcmins$ and $12\arcmins$ for SABOCA and
LABOCA respectively). The SABOCA maps are thus well matched to the SPT
resolution (FWHM $\sim$ 1\arcmin) and positional uncertainty of the SPT
source (${\rm r.m.s.}\sim10\arcsec$).

Calibration was achieved through observations of Mars, Uranus, and Neptune as
well as secondary calibrators and was found to be accurate within $10\%$ and
$25\%$ at $870$ and $350\,{\rm \mu m}$, respectively. The atmospheric attenuation was determined
via skydips every $\sim 1-2$ hours as well as from independent data from the
APEX radiometer which measures the line of sight water vapor column every
minute. Focus settings were determined typically once per night and checked
during sunrise. Pointing was checked on nearby quasars and found to be stable
within 3\arcsec\ r.m.s.

The data were reduced using the Bolometer Array analysis software (BoA)
reduction package \citep{schuller10}. The time-ordered data undergo flat
fielding, calibration, opacity correction, correlated noise removal on the full
array as well as on groups of bolometers related by the wiring and in the
electronics, flagging of unsuitable data (bad bolometers and/or data taken
outside reasonable telescope scanning velocity and acceleration limits), as well
as de-spiking. Each reduced scan was then gridded into a spatial intensity and
weighting map. Weights are calculated based on the r.m.s.\ of each time series
contributing to a certain grid point in the map. Individual maps were co-added
with inverse variance weighting. The resulting map was used in a second
iteration of the reduction to flag those parts of the time streams with
significant source signal. This guarantees that the source fluxes are not
affected by filtering and baseline subtraction. For sources that remained
undetected at $350\,{\rm \mu m}$ we flagged a region with $\rm 15\arcsecs$
radius centered on the $870\,{\rm \mu m}$ LABOCA position in the $350\,{\rm \mu
m}$ time streams.

\subsection{\textit{Spitzer}-IRAC imaging}\label{subsection:spitzerobs}
Mid-infrared {\sl Spitzer}/Infrared Array Camera (IRAC) imaging was obtained on
2009 August 02 as part of a larger {\sl Spitzer} program (PID 60194; PI Vieira)
to follow up bright submm galaxies identified in the SPT survey.  The on-target
observations consisted of $36\times 100\,{\rm s}$ and $12\times 30\,{\rm s}$ dithered
exposures at $3.6$ and $4.5\,{\rm \mu m}$, respectively.
A large dither pattern was used for the $3.6\,{\rm \mu m}$
exposures, and a medium dither pattern was used at $4.5\,{\rm \mu m}$. This scheme
was designed to provide $3.6\,{\rm \mu m}$ imaging sufficiently deep to detect even
very distant galaxies, while also providing a minimal level of sensitivity at
$4.5\,{\rm \mu m}$ to furnish infrared colors for low-redshift sources nearby.

The data were reduced following the methods described in \citet{ashby09}. The
corrected Basic Calibrated Data (cBCD) frames were modified individually to
eliminate column pull-down artifacts, and treated to remove residual images
arising from prior observations of bright sources.  The resulting pre-processed
frames were then mosaiced with standard outlier rejection techniques using
MOPEX under the control of IRACproc \citep{schuster06} to create two mosaics
covering the field. The final mosaics were generated with 0\farcs86 pixels,
i.e., pixels subtending a solid angle equal to half that of the native IRAC
pixels. The field covered with at least ten $3.6\,{\rm \mu m}$ exposures is roughly
6\arcmin\ wide, centered on the SPT position. In this work, we show only the
$3.6\,{\rm \mu m}$ images.

\subsection{Spectroscopic Redshifts}\label{subsection:specz}
To date, spectroscopic redshifts have been obtained for two sources in this sample. SPT\,0538$-$50
has a redshift of $z=2.783$, derived from ionized silicon (Si\,{\sc iv}~$\lambda$1400\AA) emission
in an optical spectrum with VLT/X-SHOOTER (284.A-5029; PI:Chapman) and carbon monoxide CO(7--6) and
CO(8--7) with APEX/Z-Spec (086.A$-$0793 and 087.A$-$0815; PI: De Breuck, and 086.F$-$9318 and
087.F$-$9320; PI: Greve). Multiple source images surrounding a low-redshift elliptical galaxy are
resolved in high-resolution $890\,{\rm \mu m}$ imaging with the Submillimeter Array (SMA)
(2009A-S076, PI: Marrone). The foreground lens has a measured
spectroscopic redshift of $z=0.443$. A paper detailing these measurements and providing detailed
characterization of this source (including lens modeling) is in preparation (Bothwell \textit{et
al.}).

SPT\,2332$-$53 has a redshift of $z=2.738$, derived from Ly$\alpha$ and
ionized carbon (C\,{\sc iv}~$\lambda$1549\AA)  in an optical spectrum
with VLT/FORS2 from program (285.A$-$5034; PI: Chapman) and carbon
monoxide CO(7--6) with APEX/Z-Spec. The redshift of the foreground
lens, which is a cluster of galaxies, is $z=0.403$. A paper providing a
full characterization of this source, including unambiguous evidence for
it being lensed by the cluster of galaxies, is in preparation (Vieira
\textit{et al.}).


\section{Results}\label{section:results}
\subsection{Submillimeter maps and source morphology}\label{subsection:maps}
\begin{figure*}
\includegraphics[width=1.0\textwidth]{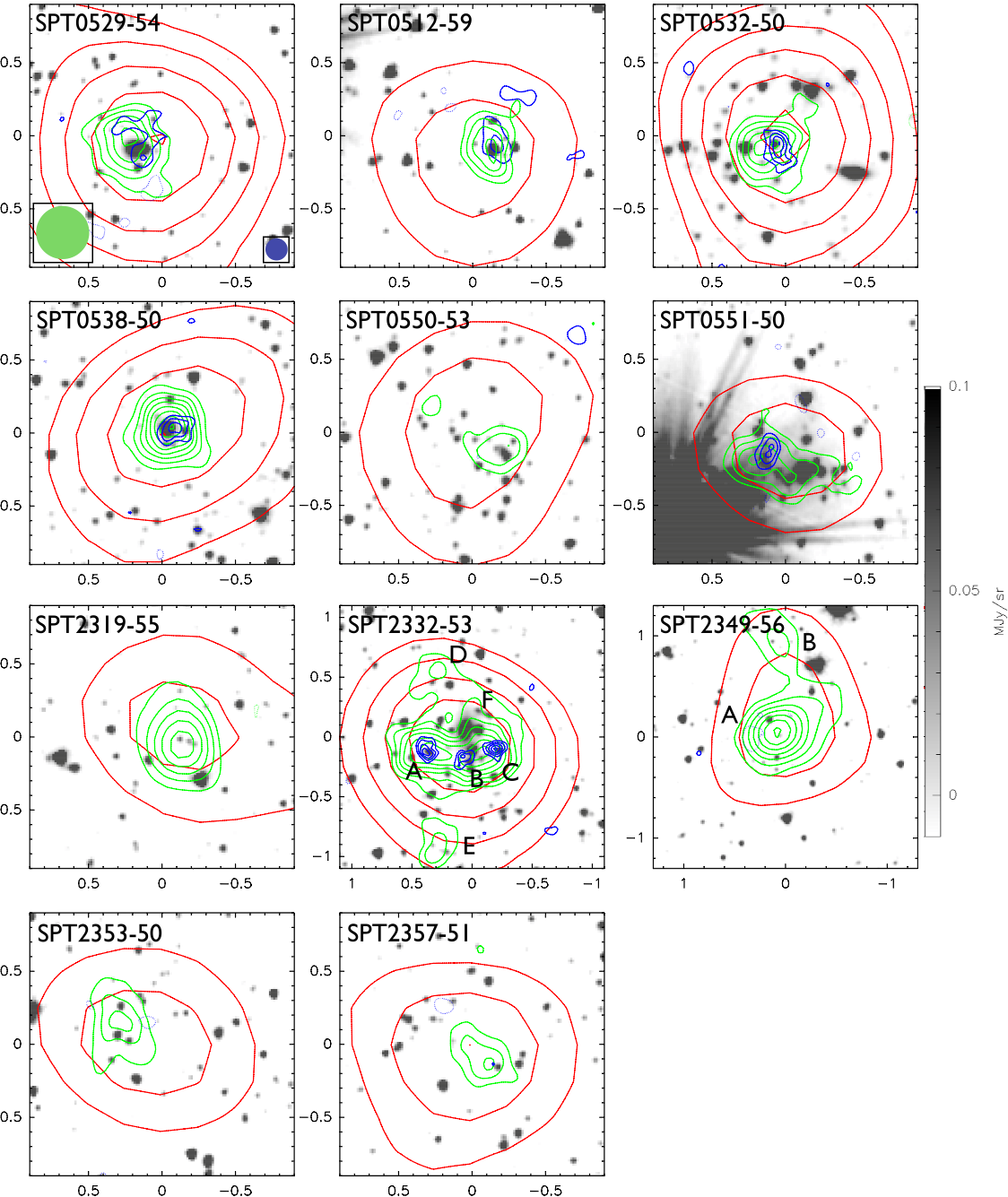}
\caption{{\it Spitzer}/IRAC $3.6\,{\rm \mu m}$ images (grey scale; $1.8'\times
1.8'$ for most, $2.2'\times 2.2'$ and $2.6'\times 2.6'$ in the cases of
SPT\,2332$-$53 and SPT\,2349$-$56, respectively) of the 11 SPT sources
presented here, with the corresponding SPT $1.4\,{\rm mm}$ $S/N$ contours
overlaid (red contours), shown at $\rm 3, 5, 7,$
etc. Also overlaid are the same $S/N$ levels (including $S/N = -3$ shown as
dashed contours) for the LABOCA $870\,{\rm \mu m}$
(green contours) and SABOCA $350\,{\rm \mu m}$ observations (blue contours).
The regions shown are centered on the SPT centroid position. The LABOCA and
SABOCA beam sizes are represented in the lower left and right corners of the
top left panel, respectively. 
}
\label{figure:saboca-laboca-maps}
\end{figure*}

Fig.\ \ref{figure:saboca-laboca-maps} shows {\it Spitzer}/IRAC $3.6\,{\rm \mu
m}$ postage-stamp images of our 11 sources, with the SPT $1.4\,{\rm
mm}$, LABOCA $870\,{\rm \mu m}$ and SABOCA $350\,{\rm \mu m}$ signal-to-noise
ratio ($S/N$) contours overlaid. 

All sources are detected at $870\,{\rm \mu m}$ at $S/N > 6$. For 6 of the 11
sources, the $S/N > 3$ in the SABOCA $350\,{\rm \mu m}$ images near the
LABOCA positions (Fig.\ \ref{figure:saboca-laboca-maps}). A further source
(SPT\,2357$-$51) is marginally detected at $350\,{\rm \mu m}$ as it shows a
$S/N \simeq 3$ emission peak at the LABOCA centroid position. Of the remaining
four sources, two (SPT\,2349$-$56 and SPT\,2319$-$55) are tentatively detected
($S/N \simeq 2.5$) in the SABOCA map, while two show no signs of a SABOCA
detection.

All sources, with the exception of SPT\,0529$-$54 and SPT\,2332$-$53, are
consistent with being point sources in both the LABOCA and SABOCA maps. Two
LABOCA sources are associated with SPT\,2349$-$56 (A and B, see Fig.\
\ref{figure:saboca-laboca-maps}), each is consistent with a point source.
Inserting point sources into the bolometer time-streams and subjecting them to
the same reduction steps as detailed in \S\ref{subsection:apexobs} reproduces
the observed maps very well.

SPT\,0529$-$54 appears to be resolved at $350\,{\rm \mu m}$, with $S/N \sim
3-4$ emission extending to the north and east. This three-component
substructure is enclosed within the area of a LABOCA beam, and the source
appears point-like at $870\,{\rm \mu m}$.

For SPT\,2332$-$53 the LABOCA $870\,{\rm \mu m}$ emission is clearly extended
in the east-west direction, spanning $ \sim 1'$. At $350\,{\rm \mu m}$ the
source resolves into three separate, unresolved blobs. We denote these sources
A, B and C, going from east to west (see Fig.\
\ref{figure:saboca-laboca-maps}). Two additional sources (D and E) are seen in
the LABOCA map to the north and south of the central east-west source.
Subtracting point sources at the A, B, and C positions in the LABOCA map
reveals an additional LABOCA source, denoted F in Fig.\
\ref{figure:saboca-laboca-maps}. While D, E and F are significant ($\gs
5-6\sigma$) at $870\,{\rm \mu m}$, no associated emission is seen in the SABOCA
map. For a detailed analysis of the properties of SPT\,2332$-$53 we refer to
Vieira \textit{et al.}, in prep.


\subsection{Millimeter and submillimeter fluxes}\label{subsection:fluxes}
The LABOCA source coordinates, the SPT flux densities at $1.4$ and $2.0\,{\rm mm}$,
and the LABOCA/SABOCA flux densities at $870$ and $350\,{\rm \mu m}$ for our
11 sources are listed in Table \ref{table:sources}. The SPT flux densities have been
determined using the method described in V10. We report the peak flux densities,
as none of the sources is resolved in the SPT maps. Submm flux density determinations
are discussed below. For all source fitting, the absolute calibration error was added in quadrature to the statistical error. 

We do not report the $2.0\,{\rm mm}$ flux of SPT\,2332$-$53 since it coincides
with a massive cluster of galaxies (see \citet{vanderlinde10}), and the SZ
decrement of the cluster at this wavelength overwhelms any emission (Vieira
\textit{et al.}, in prep). In the case of SPT\,2349$-$56, we disregard the B
component, and attribute the SPT $1.4$ and $2.0\,{\rm mm}$ fluxes to
the component A, which coincides with the SPT centroid.

The $870\,{\rm \mu m}$ LABOCA flux densities were determined by fitting a
Gaussian with $\rm FWHM=21.5\arcsecs$, the resolution of the smoothed maps, to
the sources (except for SPT\,2332$-$53, see below). A constant baseline term
was included in the fit in order to account for any background level.

The $350\,{\rm \mu m}$ SABOCA fluxes were derived by summing the flux within an
aperture corresponding to the LABOCA beam solid angle ($\rm \Omega =
1.133\times FWHM^2$), i.e.\ within a radius of $r_{\rm ap} =
\sqrt{\Omega/\pi}$. A background level was estimated from a surrounding annular
region (inner and outer radii of $r_{\rm 1} = r_{\rm ap}$ and $r_{\rm 2} =
\sqrt{2}\times r_{\rm ap}$) and removed. The aperture was centered on the
$870\,{\rm \mu m}$ LABOCA centroid. The flux uncertainties were estimated for
each source by measuring the flux within the same aperture at the same position
in 100 difference maps generated from individual observations, and measuring
the variance of the resulting distribution. For the SABOCA detections, we compared
fluxes obtained by centering the aperture on the SABOCA and LABOCA centroids,
and found the differences to be $\rm < 10\,\%$, well within the photometric
uncertainties. 

For SPT\,2332$-$53, which is resolved into three sources in the LABOCA map, we
derived the total flux within an aperture of $\rm r=44\arcsecs$. This aperture
encompasses components A, B, C, and F. The same aperture was used to measure the
corresponding total flux at $350\,{\rm \mu m}$ in the SABOCA map. The
$1.4\,{\rm mm}$ SPT flux contains the combined contribution from all four
components, and possibly also the D and E sources seen in the LABOCA maps,
because of the large SPT beam.  For subsequent photometric analysis we ignore
any contribution from the positions of D and E to the total flux because of
their large distance from the detected source. 
The D and E sources are not consistent with being part of the same lensing configuration as A, B, C, and F. The D and E sources do not have obvious IRAC counterparts. We have investigated the possibility that they are artifacts from the LABOCA data processing, but every test indicates that they are real. 

SPT\,0529$-$54 is extended in the SABOCA map, so we determined its $350\,{\rm
\mu m}$ flux using an aperture with radius $\rm 25\arcsecs$.

\subsection{Spectral energy distributions and redshifts}\label{subsection:SEDs}
The majority of our sources do not have spectroscopic redshifts, which prevents
us from determining their dust temperatures and luminosities. Our data are not
well-suited to photometric redshift measurements that rely on LIRG SED
templates \citep[e.g.][]{silva98, chary01}, as we do not have data in the MIR
and NIR where these templates have strong spectral features. The SED of dusty
galaxies is nearly featureless at the spectral resolution of bolometer cameras,
thereby hampering simple photometric redshift techniques.  The most prominent
feature, the frequency of the SED peak, cannot provide a redshift without prior
information on the dust temperature, as these quantities are degenerate through
the ratio $T_{\rm d}/(1+z)$ (e.g. \citet{blain03}).  Nonetheless, some success
in predicting spectroscopic redshifts of SMGs have been achieved by
maximum-likelihood techniques that employ libraries of SED templates of local
starbursts and ULIRGs in their analysis \citep[e.g.][]{aretxaga07}.

To overcome this degeneracy, we have developed a method that uses the
distribution of SEDs for sources of known redshift taken from the literature.
We assume a simple grey-body form for the SED to derive $T_{\rm d}$ for sources
in three samples: 1) unlensed sources at $z>1$ with $350\,{\rm \mu m}$ imaging,
2) lensed sources at $z>1$ with $350\,{\rm \mu m}$ imaging, and 3) SPT sources
with known redshifts.

For the unlensed sources, we draw our sample from: 
\begin{itemize}
\item 16 sources from \citet{kovacs06}, an $850\,{\rm \mu m}$ selected sample with 1.4 GHz
radio counterparts, optical spectroscopic redshifts, and $350\,{\rm \mu m}$
imaging from CSO/SHARC-II.\footnote{Some of these sources also have $1.1\,{\rm mm}$
photometry from CSO/BOLOCAM and/or $1.2\,{\rm mm}$ photometry from IRAM/MAMBO, which we
use, when available.}
\item 11 sources from \citet{kovacs10}, a $z\sim2$ \textit{Spitzer} IRAC and MIPS selected
sample with \textit{Spitzer}/IRS MIR spectroscopy, IRAM/MAMBO $1.2\,{\rm mm}$ imaging
and $350\,{\rm \mu m}$ imaging from CSO/SHARC-II.
\item 18 sources from \citet{chapman10}, a $1.4\,{\rm GHz}$ radio selected sample with
$850\,{\rm \mu m}$ detections, optical spectroscopic redshifts, and $250$,
$350$, and $500\,{\rm \mu m}$ imaging from \textit{Herschel}/SPIRE.
\item 13 optically and radio selected AGN from other studies \citep{benford99, beelen06, wang08,
wang10} with optical spectroscopic redshifts, mm photometry, and $350\,{\rm \mu
m}$ imaging from CSO/SHARC-II. 
\end{itemize}
For the lensed sources, we draw our sample from: 
\begin{itemize}
\item 5 well studied lensed sources from the literature, \citep[e.g.,][]{blain99c,
benford99, barvainis02} including IRAS\,F10214$+$4724, APM\,08279$+$5255, and
H\,1413$+$117 (the Cloverleaf). $350\,{\rm \mu m}$ imaging for these sources
comes from CSO/SHARC-II \citep{benford99,beelen06} and millimeter photometry
from a variety of literature sources.
\item 2 strongly lensed sources discovered behind massive galaxy clusters 
(the Eyelash and the Bullet) and imaged with SPIRE \citep{ivison10,rex10}.
\item 7 lensed sources discovered serendipitously in large extragalactic
\textit{Herschel}/SPIRE surveys \citep{negrello10,omont11, cox11,conley11}. 
\end{itemize}
For the two SPT sources, we use the measured spectroscopic redshifts described in
\S\ref{subsection:specz}. 

\begin{figure}[h]
\begin{center}
\includegraphics[width=0.48\textwidth]{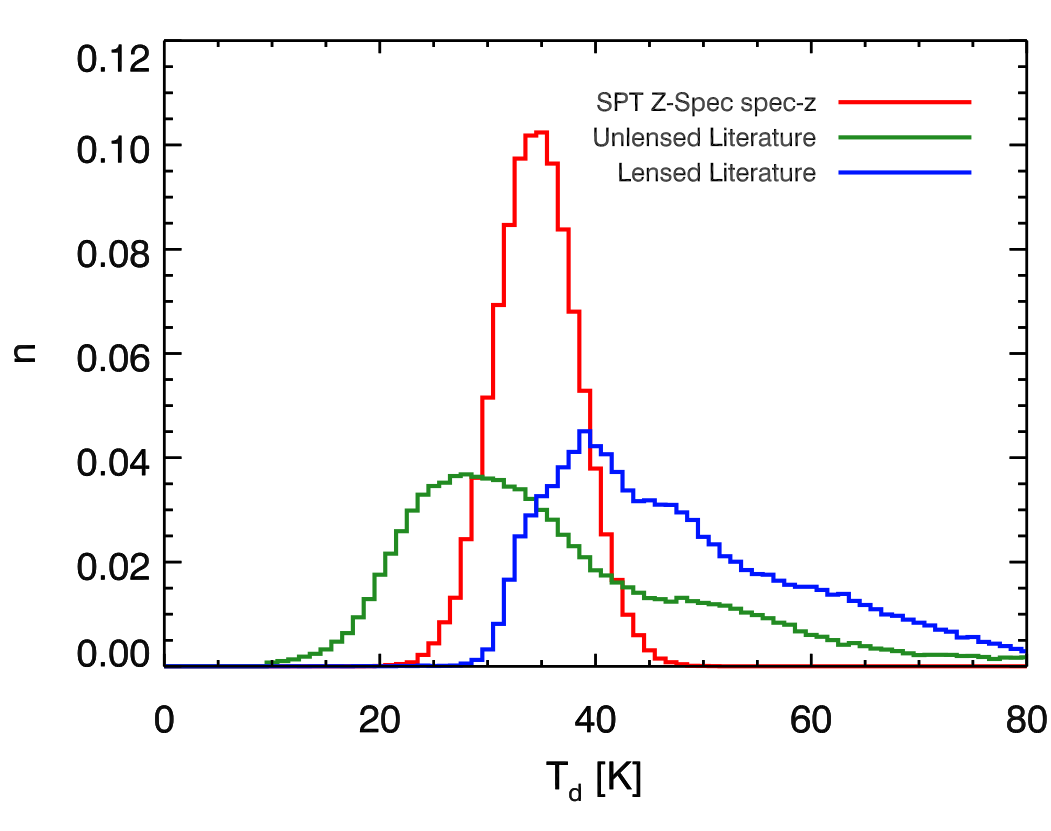}
\end{center}
\caption{A comparison of the dust temperature distributions derived from
calibration samples of: (red) two spectroscopically confirmed SPT sources,
(blue) all {\it lensed} high-$z$ sources in the literature with $350\,{\rm \mu
m}$ detections and spectroscopic redshifts, and (green) {\it unlensed}
$350\,{\rm \mu m}$-detected SMGs with spectroscopic redshifts. These
distributions are used as priors on $T_{\rm d}$ for our photometric redshift
technique, with the unlensed (green) being the primary choice for subsequent
analyses. The error bars include the statistical and absolute calibration uncertainties. 
}
\label{figure:td_hist}
\end{figure}

We fit each source in each sample with a blackbody law, modified with a
spectral emissivity that varies physically such that the dust opacity reaches
unity at frequency $\nu_{\rm c}$ \citep[e.g.,][]{blain03}:
\begin{equation}
f_{\nu}\propto [1-{\rm exp}(\nu / \nu_{\rm c})^\beta]B_{\nu}(T_{\rm d})
\end{equation}
Here, $B_{\nu}(T_{\rm d})$ is the Planck function. We fix the spectral index of the emissivity to
$\beta = 2.0$, and critical frequency to $\nu_{\rm c} \simeq 3000\,{\rm GHz}$ ($\lambda_{\rm c}
\simeq 100\,{\rm \mu m}$) following \citet{draine06}. From these fits we derive a distribution of
dust temperatures for each sample from the sum of the probability distributions of the individual
sources (Fig.\ \ref{figure:td_hist}). The SED fits exclude any data at wavelengths shorter than
$\lambda_{\rm obs}<250\,{\rm \mu m}$ (rest wavelength $\sim 50\,{\rm \mu m}$ for the highest
redshift sources) to ensure similarity between the calibration samples and the data in hand for SPT
sources in this work.  Moreover, such a constraint is appropriate to ensure the applicability of our
simple model, which cannot adequately represent emission from hot components that may become
apparent at shorter wavelengths.  We find that our simple greybody fits all sources well over this
wavelength range. We attribute the difference in the dust temperature distributions between
the lensed and unlensed sources to the former being selected near the peak of their dust emission
(e.g., by SPIRE at $350\,{\rm \mu m}$ or, in the case of the bright AGN, by IRAS at $60/100\,{\rm
\mu m}$), which will bias the selection towards hotter (and more luminous sources). The unlensed sources
were selected from either mm, radio or IRAC data, and we intentionally marginalized over these three selection
techniques to mitigate any bias.

\begin{figure*}[h]
\begin{center}
\includegraphics[width=1.0\textwidth]{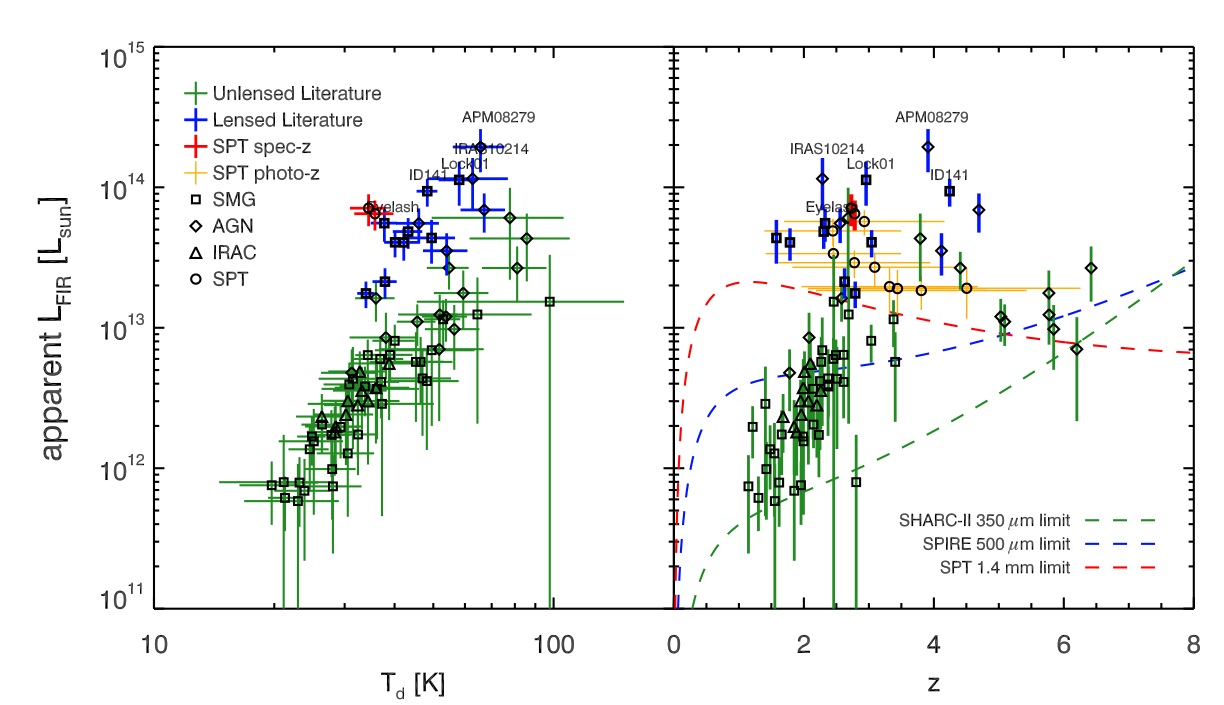}
\end{center}
\caption{\emph{\bf Left:} Apparent $L_{\rm FIR}$ vs.\ $T_{\rm d}$ for all
unlensed (\emph{green symbols}) and lensed (\emph{blue symbols}) $350\,{\rm \mu
m}$-detected SMGs and AGN (\emph{squares} and \emph{diamonds}, respectively)
with spectroscopic redshifts at $z>1$ (see \S\ref{subsection:SEDs} for
references). Apparent (i.e.\ lensed) FIR luminosities vs.\ dust temperature for
the two SPT sources with spectroscopic redshifts presented in this paper (red
symbols). For all sources, the FIR luminosities and dust temperatures were
derived by fitting a modified blackbody law with $\beta=2.0$ becoming optically
thick at $\lambda_{\rm c} < 100\,{\rm \mu m}$. \emph{\bf Right:} The FIR
luminosity as a function of redshift for the same sources. The 9 SPT sources
presented here which lack spectroscopic redshifts are also plotted (black dots
with yellow error bars) with their photometric redshift estimates and
associated uncertainties. The uncertainty on $L_{\rm FIR}$ given the assumed photometric
redshift is described in \S\ref{subsection:SEDs}, which assumes $T_{d\rm d}=34$K -- 
the median dust temperature of the unlensed population. The green, blue, and red dashed
curves show the limiting FIR luminosity as a function of redshift corresponding
to the effective SHARC-II $350\,{\rm \mu m}$ survey limit (${\rm r.m.s.}\sim
5\,{\rm mJy}$), $3\times$ the SPIRE $500\,{\rm \mu m}$ confusion limit
($30\,{\rm mJy}$), and the $3~\sigma$ SPT $2.0\,{\rm mm}$ survey limit
($3.9\,{\rm mJy}$), respectively, given the SED model of this paper and assuming a $35\,{\rm K}$
dust temperature. The SPT
survey, with its longer wavelength selection, is more sensitive to sources at
the highest redshifts ($z > 5$) than \textit{Herschel} and the large survey
area of SPT makes it sensitive to sources that are amongst the most rare and most highly magnified. 
}
\label{figure:lum}
\end{figure*}

The distribution of $T_{\rm d}$, apparent $L_{\rm FIR}$, and $z$ for the three
samples can be seen in Fig.\ \ref{figure:lum}.  The lensed sources have the
highest apparent luminosity at a given $T_{\rm d}$, as expected from the
gravitational magnification, which increases the solid angle they subtend. The
lensed sources appear to be offset in the $L_{\rm FIR}$ vs. $T_{\rm d}$ space,
indicative of them being randomly sampled from the background and 
made more luminous by gravitational magnification. 
The lensed sources pulled from the literature are significantly hotter than the unlensed population. 
This is presumably due to selection effects. 

The detection limits for the types of observations used in the three samples
are also shown in the right panel of Fig.\ \ref{figure:lum}, for the SED model
employed in this work and assuming a dust temperature of $35\,{\rm K}$. The
apparent luminosity threshold decreases at high redshift for SPT selection, owing to the
steep rise in the submillimeter SED of these objects for $\beta=2$. 

\begin{figure*}[h]
\begin{center}
\includegraphics[width=1.0\textwidth]{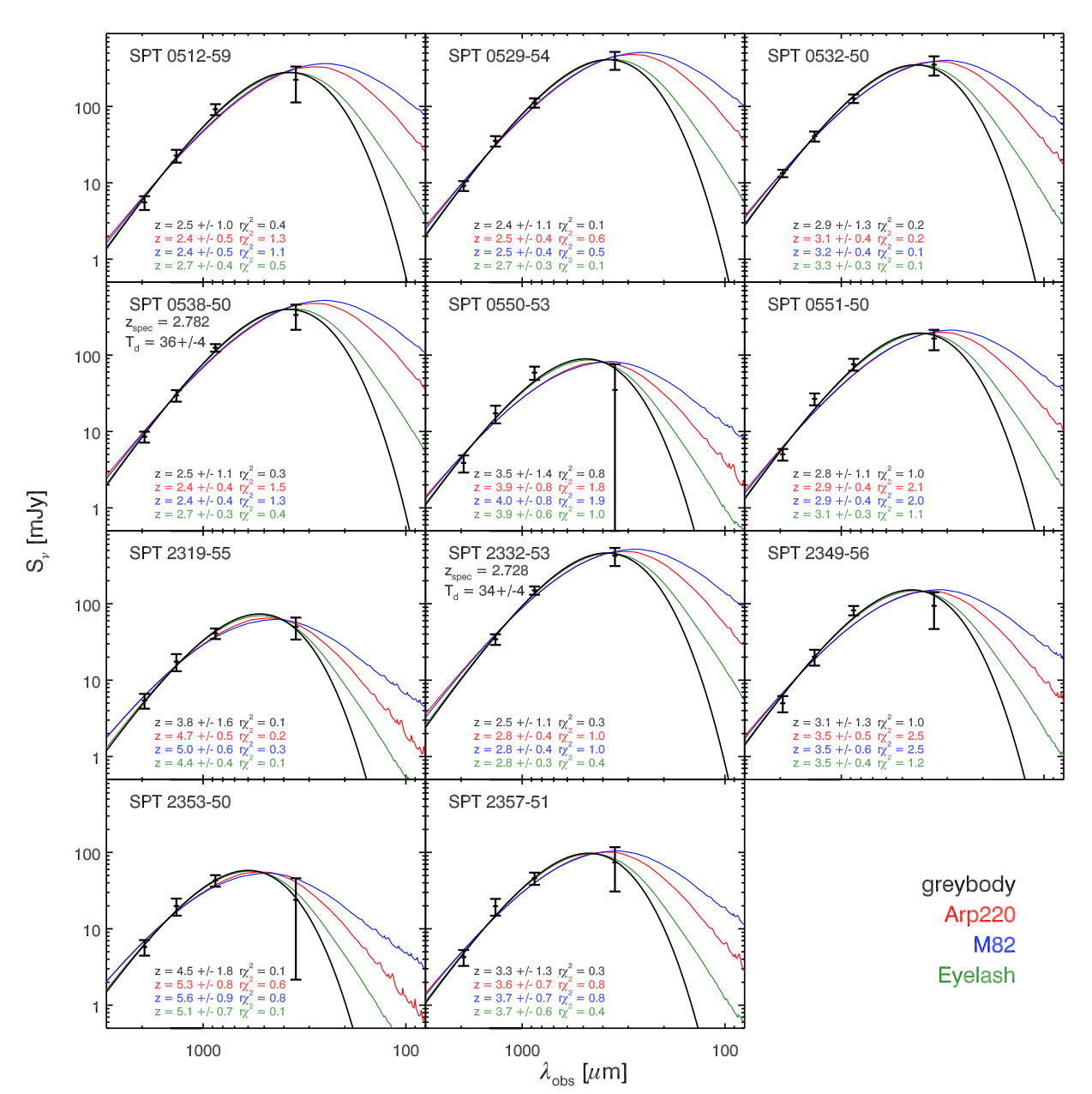}
\end{center}
\caption{Observed submm/mm SEDs for the 11 sources presented here.  The black
symbols are SABOCA ($350\,{\rm \mu m}$), LABOCA ($870\,{\rm \mu m}$) data, and
SPT photometry ($1.4$ and $2.0\,{\rm mm}$). The black curves are modified
blackbody fits with fixed spectral index ($\beta=2.0$), and $T_{\rm d}=34$K taken from the median value of the unlensed sample of SMGs. 
Spectroscopic redshifts break the $T_{\rm
d}/(1+z)$ degeneracy \citep{blain03}, and in those cases we quote the best-fit
dust temperature. The red, blue, and green curves represent the best-fit SEDs
to the data based on the Arp\,220, M\,82, and Eyelash SED templates,
respectively, (see \S\ref{subsection:SEDs}), where the absolute scaling and the
redshift has been allowed to vary (the latter is given in each panel).  The
greybody fit in black is the simplest curve which fits the data.  Without
additional photometry shortward of $350\,{\rm \mu m}$ it is difficult to
discriminate between different SED templates such as Arp\,220 or M\,82. 
}
\label{figure:SEDs}
\end{figure*}

The observed submm/mm SEDs of the SPT sources are shown in Fig.\
\ref{figure:SEDs}. In all cases, the $350\,{\rm \mu m}$ data point falls below
the power-law extrapolation of the longer-wavelength data, giving the
photometric redshift technique a spectral feature from which to impose its
constraint. To infer a photometric redshift for an SPT source, we compare our
data to the modified blackbody SED model described above.  We randomly draw a
value of $T_{\rm d}$ from one of our distributions and fit the redshift for the
SED described by this temperature.  Repeating this $10^4$ times, we generate a
probability distribution for the source redshift that marginalizes over the
$T_{\rm d}$ prior, and adopt the median and standard deviation of this
distribution as the estimate of the photometric redshift and its uncertainty.
The unlensed sample has a median dust temperature of $\sim 34\,{\rm K}$, while
the lensed sample has a median dust temperature of $\sim 46\,{\rm K}$. We adopt
the distribution from the unlensed sources throughout this paper under
the assumption that lensing is randomly sampling the underlying unlensed
population of sources. This is a conservative choice, since of the three distributions, 
it results in the lowest
redshifts because of the $T_{\rm d}/(1+z)$ degeneracy. Also, the dust temperatures derived
for the two sources with spectroscopic redshifts fall in the middle of the unlensed distribution.
The photometric redshifts derived from this analysis are indicated in Fig.\ \ref{figure:SEDs}
and in Table \ref{table:results}.

As a point of comparison, we also fit each SPT source with redshifted SED
templates of Arp\,220, M\,82 \citep{silva98}, and the Eyelash \citep{ivison10}.
The Eyelash SED is modeled using a two-component dust model comprising two
modified blackbodies with $T_{\rm d} = 30$ and $60\,{\rm K}$ and $\beta=2$, as
described in \citet{ivison10}. We adopt the redshift for which a given
template provided the best fit to the data (global $\chi_{\nu}^2$ minimum) as
the best estimate photometric redshift (for that template).  The photometric
redshifts and uncertainties obtained from these three SED templates are shown
in Fig.\ \ref{figure:SEDs}. Without additional photometry shortward of
$350\,{\rm \mu m}$ it is difficult to discriminate between different SED
templates such as Arp\,220 or M\,82. These galaxy template redshifts are
consistent with the temperature-marginalized greybody values.

For the two sources with spectroscopic redshifts (SPT\,2332$-$53 and
SPT\,0538$-$50), we can also directly check the photometric redshift estimates.
In both cases we recover the true redshift to within the statistical
uncertainties.

\subsection{FIR luminosities, star formation rates, and dust masses}\label{subsection:mBB}
Adopting the photometric redshifts and SED fits described above in \S\ref{subsection:SEDs}, we can
derive far-infrared luminosities and star formation rates, and dust masses. The fits are shown as
black curves in Fig.\ \ref{figure:SEDs}, and are seen to provide a good match to the data. For the
two sources with spectroscopic redshifts, we can break the $T_{\rm d}/(1+z)$ degeneracy, and derive
the FIR luminosities, star formation rates, and dust temperatures, which are listed in Table
\ref{table:results}. For the remaining sources, we derive FIR luminosities, star formation rates,
and dust masses by fixing the dust temperature at the median of the distribution for the unlensed
sources ($T_{\rm d}=34$K).

We determine dust masses according to
\begin{equation}
M_{\rm d} = \mu^{-1}\frac{D_{\rm L}^2 S_{\nu_{\rm o}} }{(1+z) \kappa_{\nu_{\rm r}}} [B_{\nu_{\rm r}}(T_{\rm d}) - B_{\nu_{\rm r}}(T_{\rm CMB}
(z))]^{-1},
\end{equation}
where $S_{\nu_{\rm o}}$ is the flux density at the observed
frequency $\nu_{\rm o} = \nu_{\rm r} (1+z)^{-1}$ (which we here set to $345\,{\rm GHz}$,
the rough central frequency of the LABOCA bandpass). $D_{\rm L}$ is the
luminosity distance, $\mu$ is the magnification factor, and
$T_{\rm CMB}(z)$ is the cosmic microwave background temperature at 
redshift $z$ (which, in principle, has to be included, although it only
changes the dust mass at the $\sim 2\%$ level -- \citet{papadopoulos00}). We adopt 
$\kappa_{\nu_{\rm r}}/{\rm m^2\,kg^{-1}} = 0.045 \times \left (\nu_{\rm r}/250\,{\rm GHz} \right)^{\beta}$
\citep{hildebrand83,krugel94}, where $\beta=2.0$ is the adopted dust emissivity
index (see \S\ref{subsection:SEDs}). The dust masses (before correcting downward for the unknown magnification) are in the
range $M_{\rm d} \sim 0.4-1.4\times 10^{10}\,\Msolar$ (Table
\ref{table:results}). We note that significant uncertainties are associated with these estimates
due to the uncertainties in the dust emissivity.

The apparent FIR luminosity is determined from the modified blackbody SEDs,
assuming $T_{\rm d}=34\,{\rm K}$, as described in
\S\ref{subsection:SEDs}.  The SED is integrated over the rest wavelength range
$8-1000\,{\rm \mu m}$.  This yields apparent FIR luminosities in range $L_{\rm FIR} \sim
1.8-6.5\times 10^{13}\,\Lsolar$ (Table \ref{table:quantities}). To
derive total star formation rates, we use the following conversion
from \citet{murphy11}: 
\begin{equation} 
\frac{SFR}{\rm \Msolar\,yr^{-1}} = 1.49\times 10^{-10} \mu^{-1} \frac{L_{\rm FIR}[{8-1000\,{\rm \mu m}}]}{\Lsolar}, 
\end{equation} 
The resulting apparent star formation rates are in the range $\sim 2.7\times10^3-9.7\times10^3\,{\rm
\Msolar\,yr^{-1}}$ (Table \ref{table:quantities}).  No correction has been applied to
$L_\mathrm{FIR}$ or the star formation rate for the unknown gravitational magnification factor and
we use the factor of $\mu$ to indicate that these values are upper limits.  Even if the lensing
correction was known, we have assumed that AGN do not contribute significantly to the FIR
luminosity, and the star formation rates would therefore still be upper limits. 

Assuming a gas-to-dust mass ratio of $\sim 100$, consistent with what is found
for normal SMGs (e.g.,\ \citealt{santini10}), we infer total apparent gas
masses in the range $M_{\rm gas}\sim 0.4-1.4\times 10^{12}\,\Msolar$, and gas-depletion time
scales of $t_{\rm depl}\sim M_{\rm gas}/SFR \sim 100-220\,{\rm Myr}$.  This is
consistent with the lower-limit gas-depletion time scales ($t_{\rm depl} >
40\,{\rm Myr}$) inferred from CO observations of normal SMGs \citep{greve05}.
Note that the gas-depletion time scale estimate is independent of the lensing
amplification factor as it enters in both the estimate of total gas mass and
the star formation rate.


\section{Discussion}\label{section:discussion}
\subsection{FIR luminosities and dust temperatures}
In this section, we compare the derived properties of our SPT sources with those
of other ultra-bright SMGs from the literature with $350\,{\rm \mu m}$ measurements 
and spectroscopic redshifts of $z>1$. 
This comparison is summarized in Fig.\ \ref{figure:lum}, which shows
$L_{\rm FIR}$ vs.\ $z$ and $T_{\rm d}$ vs.\ $L_{\rm FIR}$ for the SPT sources
and other ultra-bright SMGs, normal SMGs, and AGN.

As can be seen in Fig.\ \ref{figure:lum}, where we plot $L_{\rm FIR}$ vs.\
$T_{\rm d}$, the two SPT sources with spectroscopic redshifts (and thus
well-determined dust temperatures) are significantly cooler than other lensed,
ultra-bright sources with similar FIR luminosities.  The dust temperatures
(determined using the SED fit described in \S\ref{subsection:SEDs}) of the
strongly lensed sources from the literature are in the range $T_{\rm
d}=34-67\,{\rm K}$  with a median of $46$~K, while the SPT dust temperatures
have a mean $T_{\rm d}=35\,{\rm K}$, see Table \ref{table:results} and Fig.\
\ref{figure:td_hist}. The fact that the SPT sources appear to have lower dust
temperatures than other ultra-bright SMGs with similar FIR luminosities is likely
due to the shorter wavelength ($<1\,{\rm mm}$) selection of the latter,
which will be biased towards warmer temperatures. 
Also, the ultra-bright SMGs (and AGN) from the literature were selected based on their
extreme brightness, which probably implies a preference towards warmer dust temperatures.
It is also possible that strong lensing preferentially selects towards compact and warmer objects.
Finally, differential
magnification may have an effect on the measured source temperatures
\citep[e.g.][]{blain99c}, but it is beyond the scope of this paper to address
this issue. 

For SPT sources with only photometric redshifts, we infer apparent FIR
luminosities (integrated from $8-1000\,{\rm \mu m}$) in the range $\sim 2 -
6\times 10^{13}\,\Lsolar$. In Fig.\ \ref{figure:lum} we plot the FIR
luminosities as a function of redshift, along with those of the above mentioned
lensed and unlensed SMGs from the literature.  The SPT sources have apparent
FIR luminosities comparable to those of other lensed, ultra-bright SMGs, such
as the SPIRE sources with spectroscopic redshifts, which have apparent FIR
luminosities in the range $3.9-7.8\times 10^{13}\,\Lsolar$.  In comparison,
normal SMGs have an average intrinsic FIR luminosity of $2.5\times
10^{12}\,\Lsolar$ (with a range $\sim 10^{11} - 3\times 10^{13}\,\Lsolar$ --
\citet{kovacs06}). Thus, the apparent luminosities of ultra-bright
SMGs (including those discovered by SPT) are typically an order of magnitude higher than those of the
normal SMGs. To a large extent, this difference is expected to reflect the
lensing amplification of the ultra-bright SMGs. In the case of the Eyelash,
where the amplification factor is well-known ($\mu =32$), the intrinsic FIR
luminosity obtained after correcting for lensing is $2.3\times
10^{12}\,\Lsolar$ \citep{ivison10}, which is typical of normal SMGs.

\subsection{SPT sources are strongly lensed}\label{subsection:lensing}
High-resolution follow-up observations of extremely bright SMGs in the literature
have shown the majority to be strongly lensed, dusty galaxies \citep{swinbank10,fu12}. It seems
reasonable to assume that the SPT sources, given their similarity to the literature sources, 
are also highly lensed, although, we cannot rule out that some fraction of them are intrinsically 
extremely luminous.

For some of the SPT sources, there is indeed unequivocal evidence that they are
strongly lensed. In the case of SPT\,2332$-$53, the LABOCA source is resolved
into three distinct SABOCA sources -- a morphology which strongly suggests
we are dealing with a single source undergoing strong gravitational lensing.
This is confirmed in optical/near-IR imaging of this source, where the SABOCA
sources are clearly seen to line up with a giant arc (Vieira \textit{et al.},
in prep.). Similarly in SPT\,0529$-$54, the LABOCA source is resolved into an
extended SABOCA source with multiple peaks, again indicative of strong lensing.
Finally, in at least one other case (SPT\,0538$-$50), we see structure in the
submm emission consistent with gravitational lensing (Bothwell \textit{et al.},
in prep.). A lack of high resolution (sub)mm imaging prevents us from
constructing accurate lens models and deriving the intrinsic properties of the
remainder of the sources. The mass scale of the haloes responsible for lensing
the SPT population (i.e.\ galaxy-scale lensing, cluster lensing, etc.) is
still poorly determined, although theoretical modeling indicates that the
majority of the sources should be lensed by massive elliptical galaxies at
$z\sim1$ \citep{hezaveh11}. Future publications will address the lensing
statistics for this sample. 

\begin{figure}[t]
\begin{center}
\includegraphics[width=0.48\textwidth]{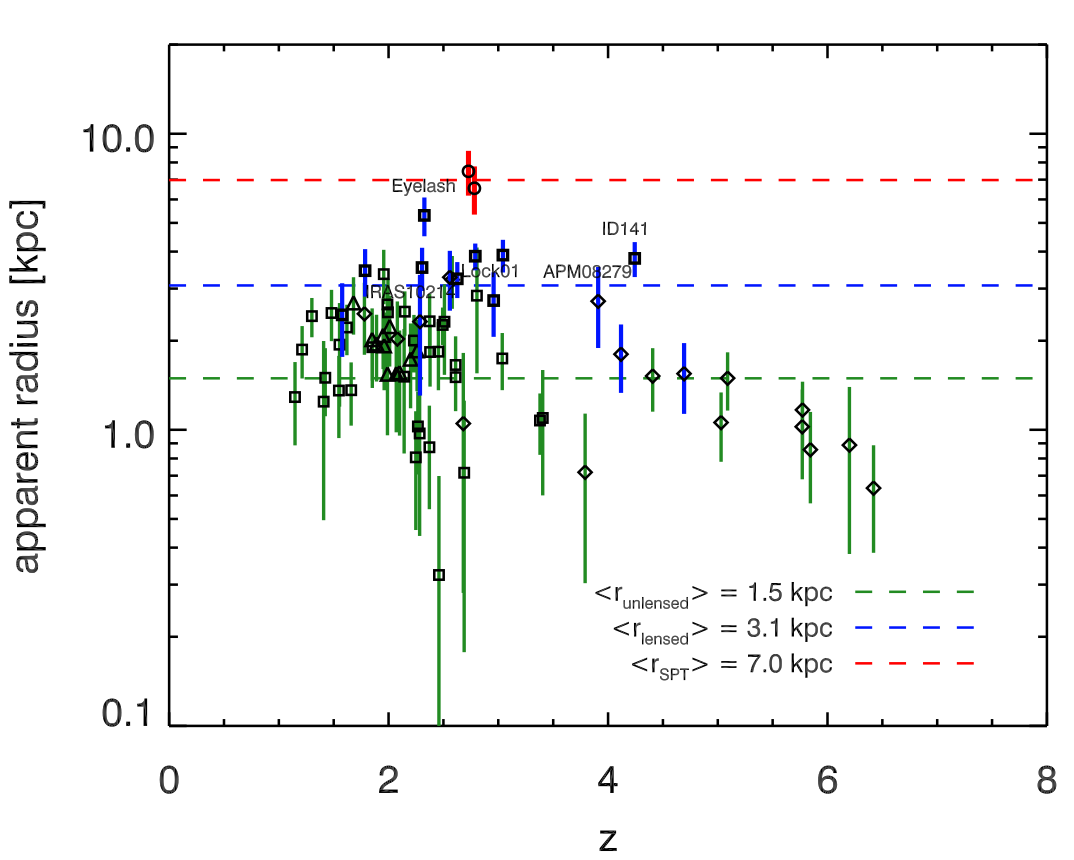}
\end{center}
\caption{The apparent radius (parametrized as $\propto \left ( L_{\rm FIR}/T_{\rm d}^4\right )^{1/2}$
in units of kpc) of the two SPT sources with spectroscopic redshifts (red
symbols). Other strongly lensed sources (SMGs and dusty AGN -- blue symbols)
from the literature are also shown, as are the normal, unlensed SMGs (green
symbols). In the text we argue that the larger apparent radius of the SPT sources is due to gravitational lensing. 
}
\label{figure:magnification}
\end{figure}

The SED fitting performed in \S\ref{subsection:SEDs} provides a way to
estimate the average magnification of the SPT sources. The luminosity
and temperature of the sources are related by a version of the Stefan-Boltzmann
law, modified to account for the frequency-dependent emissivity of the sources,
so that  $L_{\rm FIR} = 4 \pi R^2 \sigma T_{\rm d}^4$, where we assume that the
SPT sources emit as modified blackbodies isotropically from a surface of radius
$R$. The effective Stefan-Boltzmann constant, $\sigma_{\rm eff}$, is determined
for our model by setting $\sigma_{\rm eff}/\sigma$ equal to the ratio of the
integral of the modified blackbody over a perfect Planck blackbody.  For our
model, with the opacity reaching unity at $\lambda_c=100\,{\rm \mu m}$ and
$T_{\rm d}=34-36\,{\rm K}$ (as for the two SPT SMGs with spectroscopic
redshifts), $\sigma_{\rm eff}/\sigma$ varies from 0.51 to 0.54.  Using this
relation, we solve for the apparent effective radii obtained for the SPT
sources, and also for the lensed and unlensed SMGs and dusty AGN from the
literature\footnote{For simplicity we have assumed that the modification of the
Stefan-Boltzmann law is the same for all sources, given by the value for
$T_{\rm d}=34\,{\rm K}$, the median for the unlensed population. It varies by a
factor of $\pm$50\% over the $20-50\,{\rm K}$ range of $T_{\rm d}$ enclosing
the bulk of the unlensed SMGs and asymptotes to unity at large $T_{\rm d}$, but
factors of order unity are unimportant for this argument.} Assuming that the
intrinsic sizes and dust temperatures of the SPT sources are, on average,
similar to those of the unlensed SMGs, we can ascribe the differences in the
apparent radii to different average magnification factors. From Fig.\
\ref{figure:magnification} we find an average size ratio of $\langle R_{\rm
SPT}\rangle/\langle R_{\rm SMG, unlensed}\rangle \simeq 4.7$, which translates
into a magnification ratio of $\langle{\mu_{\rm SPT}}\rangle \sim 22$.
We stress, that while the apparent size estimates will depend on the optically thick
transition-wavelength, $\lambda_{\rm c}$, all sources are fitted with the same SED model, and so
changing $\lambda_{\rm c}$ will not alter the relative trend seen.

For the SPT sources with photometric redshifts, the implied lensing magnification strongly depends on the assumed dust temperature. 
Sampling the distribution of temperatures associated with the unlensed catalog, we find a broad distribution of magnifications for the SPT photometric sources with 67\% of the total samples indicating lensing magnification $\mu>5$. 
Fixing the dust temperature for all sources to be $33{\rm K} <T_{\rm d}<35$K, as was found for the two SPT sources with spectroscopic redshifts, we find the mean magnification of the SPT sources with photometric redshifts to be $\overline{\mu} = 11$. In order to make the SPT sample of photometric redshift sources consistent with no lensing $\overline{\mu} \le 1$, it is necessary for the mean dust temperature to be greater than $\overline{T_{\rm d}} > 80$K.

\begin{figure}[h]
\begin{center}
\includegraphics[width=0.48\textwidth]{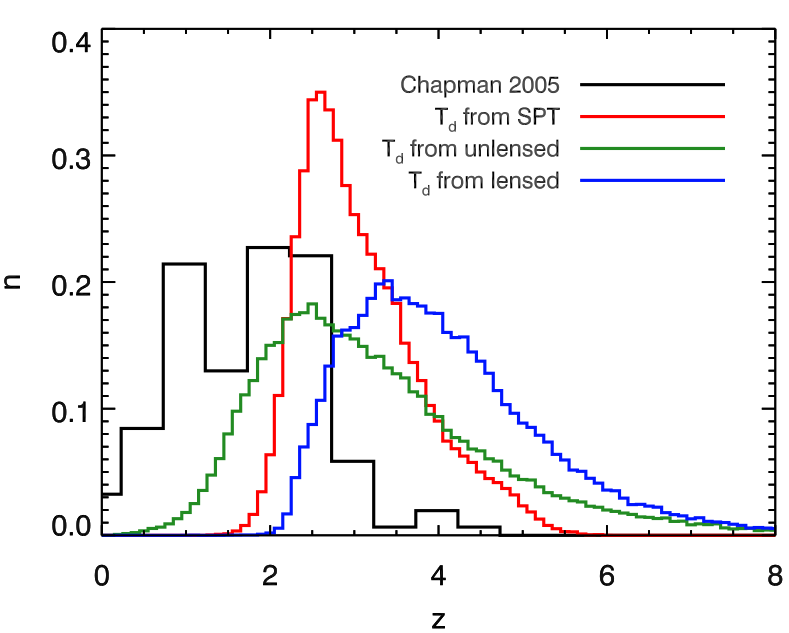}
\end{center}
\caption{A comparison of our estimated redshift distributions of the SPT
sources assuming a $T_{\rm d}$ distribution derived from: (red) the two spectroscopically
confirmed SPT sources, (blue) all {\it lensed} high-$z$ sources in
the literature with $350\,{\rm \mu m}$ detections and spectroscopic redshifts,
and (green) {\it unlensed} $350\,{\rm \mu m}$-detected SMGs with
spectroscopic redshifts. The black solid histogram shows the normalized
spectroscopic redshift distribution of normal unlensed $850\,{\rm \mu
m}$-selected SMGs based on an updated version of the \citet{chapman05} sample
\citep{banerji11}, 
and including the five spectroscopically confirmed normal $z > 4$ SMGs. 
Note that the black histogram are for individual spectroscopic redshifts, while the 
colored lines correspond to the probability distributions of photometric redshifts for the SPT sample. 
All curves have been normalized such that the integral is equal to unity.
}
\label{figure:Nz}
\end{figure}

\subsection{The redshifts of ultra-bright SMGs}\label{subsection:Nz}
From the 11 SPT sources presented in this paper, we find a median redshift of
$\overline{z}=3.0$ with a 68\% range of $z=2.0-4.6$ using the modified blackbody
fits (see \S\ref{subsection:SEDs}). If we instead use the Arp\,220, M\,82, and
Eyelash SED templates, we find in all cases median redshifts of
$\overline{z}=3.1-3.2$. We emphasize that while the photometric redshifts
derived in \S\ref{subsection:SEDs} should not be considered more reliable than
within $\Delta z \sim 1$ on a source-by-source basis, our analysis does
indicate than when taken as a sample (which includes two spectroscopic
redshifts) the SPT sources do have a very high average redshift. The use of the
unlensed sample of SMGs as the $T_{\rm d}$ prior is conservative in that it
results in lower redshifts than would be found by using the lensed sample. The
agreement between the modified black-body and template-derived redshift
distributions, and the comparison to the two so SPT sources with spectroscopic 
redshifts suggests that this assumption is reasonable. 

The redshift distribution for the SPT sources is shown in Fig.\ \ref{figure:Nz}
for all three choices of the calibration sample for the modified blackbody
fitting technique. The redshifts for a sample of normal, radio-identified,
$850\,{\rm \mu m}$-selected SMGs are also shown ($\overline{z}=2.2$, \citealt{chapman05}), 
where we have updated the distribution to include five
additional unlensed SMGs spectroscopically confirmed to reside in the redshift
range $z=4.05-5.3$ \citep{capak08,capak11,knudsen08,coppin10} as well as
newly obtained redshifts by \citet{banerji11} of SMGs
in the spectroscopic 'redshift desert' ($z=0.8-1.5$). If the SPT
sources have similar dust temperatures as the unlensed SMG population
at $z>1$ with spectroscopic redshifts, then the FIR colors imply that the
SPT sources are at systematically higher redshifts.  For all choices of
calibration sample (which provide the $T_{\rm d}$ prior), the SPT sources are
at higher median redshift than the \citet{chapman05} sources.

The SPT sources are hypothesized to be a sample of gravitationally magnified
members of the normal SMG population, which suggests that they should be drawn
from the same redshift distribution. The difference in their observed redshift
distributions is likely due to a combination of selection effects. First, the
\citet{chapman05} sample is strongly selected against high redshift objects by
virtue of the reliance on radio detections to inform spectroscopic followup.
Some correction is provided by the addition of newer sources found to be at
higher redshift, but the selection remains important.  For the SPT sample, we
removed any source which is detected in the IRAS-FSC, thereby removing low
redshift unlensed interlopers and possibly a small number of lensed sources at
$z\sim1$. The longer-wavelength selection of the SPT sample also admits very
high-redshift sources that would be harder to detect at $850\,{\rm \mu m}$ as
the SED peak approaches the bolometer passband. Finally, the strong lensing
condition itself is expected to introduce a bias towards higher redshift
objects.  Theoretical work predicts that gravitational magnification by
clusters and galaxies shifts the redshift distribution of an observed source
population toward higher redshifts than an equivalent field-selected sample
\citep{broadhurst95,blain96,negrello07,jain11}.  

\section{Summary}
We have presented SABOCA $350\,{\rm \mu m}$ and LABOCA $870\,{\rm \mu m}$ maps
and fluxes of 11 ultra-bright SMGs from the SPT survey, resulting in the first
FIR/submm SEDs of SPT sources. Employing a novel method, which makes use of a
compilation of SEDs of 58 sources with spectroscopic redshifts at $z > 1$ and with
$350\,{\rm \mu m}$ photometry, we have derived photometric redshift estimates
for our sources. In the two cases where we have spectroscopic redshifts, they
agree well with the photometric redshifts from this analysis. The SPT redshift
distribution, with a mean of $\overline{z}=3.0$ and an 68\% confidence region
$z=2.0-4.6$, is found to be skewed towards significantly higher redshifts than
the distribution observed for the normal un-lensed SMG population. We
argue that the higher mean redshift of the SPT sources is due to the longer
selection wavelength and the lensing selection-criterion. 

Using the median dust temperature measured from the unlensed population of sources from the literature, 
we fit modified blackbody laws to the observed SPT SEDs and estimated
their apparent (i.e., lensing-uncorrected) FIR luminosities ($L_{\rm FIR} \sim
3.7\times 10^{13}\,\Lsolar$), star formation rates ($SFR \gs
2500\,{\rm\Msolar\,yr^{-1}}$), and dust masses ($M_{\rm d} \sim
0.4-1.9\times 10^{10}\,\Msolar$). Based on the two SPT sources with
spectroscopic redshifts, we derive dust temperatures ($\sim 35\,{\rm K}$)
similar to that of normal SMGs. From the same two sources we estimate a value
of $\sim 22$ for the gravitational magnification factor. Such a large
magnification factor is in line with tentative lens models (Vieira \textit{et
al.} in prep, Bothwell \textit{et al.} in prep) and theoretical modeling
\citep{hezaveh11}. Once corrected for lensing, our findings, as well as those
of others \citep[e.g.][]{conley11}, suggest that the intrinsic properties of
the ultra-bright SMGs will fall within the range of normal SMGs. Based on
theoretical models and the measured projected area, it appears that the wide field
coverage of the SPT finds the rarest, brightest objects, including those 
with the highest magnification relative to strongly lensed SMGs and AGN
found in the literature. 

Due to the extraordinary boosting of the flux and angular size caused by the
gravitational lensing, ultra-bright SMGs are ideal targets for high-resolution
follow-up observations with the Atacama Large Millimeter Array (ALMA),
\textit{Herschel Space Observatory}, and the \textit{Hubble Space Telescope}.
These observatories can access the dust properties, gas kinematics, and star
formation conditions on sub-kpc scales in these systems. 
These observations were based on sources selected from 200 deg$^2$. 
SPT has now surveyed 2500 deg$^2$ and an extensive multi-wavelength campaign is on-going. 

\acknowledgments
The authors would like to thank A.\ Blain and N.\ Scoville for stimulating
discussions and guidance.  We are grateful to the competent staff at the APEX
base-camp in Sequitor, Chile. The South Pole Telescope is supported by the
National Science Foundation through grants ANT-0638937 and ANT-0130612.
Partial support is also provided by the NSF Physics Frontier Center grant
PHY-0114422 to the Kavli Institute of Cosmological Physics at the University of
Chicago, the Kavli Foundation and the Gordon and Betty Moore Foundation. TRG
acknowledges support from the UK Science and Technologies Facilities Council, as well as
from IDA and DARK. 
The McGill group acknowledges funding from the National Sciences and
Engineering Research Council of Canada, Canada Research Chairs program, and the
Canadian Institute for Advanced Research. M.D.\ acknowledges support from an
Alfred P.\ Sloan Research Fellowship.  AAS acknowledges support by the Becker
Fund of the Smithsonian Institution.  This research has made use of the
NASA/IPAC Extragalactic Database (NED) which is operated by the Jet Propulsion
Laboratory, California Institute of Technology, under contract with the
National Aeronautics and Space Administration. Finally, we are grateful to the
directors of ESO and the SMA for granting us director's discretionary time.



\bigskip

{\it Facilities:} \facility{APEX (SABOCA, LABOCA, Z-Spec)}, \facility{\textit{Spitzer}}, \facility{SPT}.

\bibliographystyle{apj}
\bibliography{spt_smg}

\clearpage

\clearpage

\begin{deluxetable*}{lllllll}
\tablecolumns{7}
\tablewidth{0pt}
\tablecaption{Flux Densities of SPT sources\label{table:sources}}
\tablehead{ID                  & R.A.           & Dec.              & $\rm S_{350\mu m}$            & $\rm S_{870\mu m}$           & $\rm S_{1.4mm}$    & $\rm 
S_{2.0mm}$ \\
               	               &                &                   & mJy                           &  mJy                         & mJy                    & mJy }
\startdata
SPT-S~J051259$-$5935.6         & 05:12:57.8     & $-$59:35:39.6     & $223\pm 95$					& $92\pm 12$	               & $22.7\pm 3.9$          & $5.5\pm 1.1$        \\
SPT-S~J052903$-$5436.6         & 05:29:03.0     & $-$54:36:33.3     & $409\pm 35$\tablenotemark{a}  & $112\pm 11$                  & $35.4\pm 4.8$          & $9.2\pm 1.3$        \\
SPT-S~J053250$-$5047.1         & 05:32:50.9     & $-$50:47:10.0     & $353\pm 47$	                & $127\pm 10$                  & $40.8\pm 5.2$          & $13.4\pm 1.4$       \\
SPT-S~J053816$-$5030.8         & 05:38:16.5     & $-$50:30:52.5     & $336\pm 88$					& $125\pm 7$                   & $29.7\pm 4.6$  	    & $8.5\pm 1.4$        \\
SPT-S~J055002$-$5356.6         & 05:50:00.5     & $-$53:56:41.2     & $35\pm 40$\tablenotemark{b}	& $59\pm 10$	               & $17.3\pm 4.3$          & $3.9\pm 1.0$        \\
SPT-S~J055138$-$5057.9         & 05:51:39.2     & $-$50:57:59.4     & $165\pm 27$					& $76\pm 11$	               & $26.7\pm 4.2$          & $5.0\pm 0.9$        \\
SPT-S~J231921$-$5557.9         & 23:19:21.5     & $-$55:57:57.6     & $50\pm 10$\tablenotemark{b}	& $41\pm 5$	  	               & $17.5\pm 4.2$          & $5.4\pm 1.2$        \\
SPT-S~J233227$-$5358.5         & 23:32:26.5     & $-$53:58:39.8     & $425\pm 39$\tablenotemark{c}  & $150\pm 11$\tablenotemark{c} & $34.4\pm 4.7$          & . . .\tablenotemark{d}               \\
. . . . . . . -A               & 23:32:29.7     & $-$53:58:38.4     & $148\pm 21$\tablenotemark{e}  & $51\pm 7$                    & . . .                  & . . .               \\
. . . . . . . -B               & 23:32:27.6     & $-$53:58:42.8     & $92\pm 21$\tablenotemark{e} 	& $38\pm 8$                    & . . .                  & . . .               \\
. . . . . . . -C               & 23:32:25.8     & $-$53:58:38.0     & $137\pm 21$\tablenotemark{}  & $43\pm 8$                    & . . .                  & . . .               \\
. . . . . . . -D               & 23:32:29.0     & $-$53:57:59.0     & $65\pm 21$\tablenotemark{b}   & $22\pm 4$                    & . . .                  & . . .               \\
. . . . . . . -E               & 23:32:29.4     & $-$53:59:31.0     & $39\pm 21$\tablenotemark{b}   & $21\pm 5$                    & . . .                  & . . .               \\
. . . . . . . -F               & 23:32:27.1     & $-$56:58:17.0     & $17\pm 21$\tablenotemark{b}	& $18\pm 4$                    & . . .                  & . . .               \\
SPT-S~J234942$-$5638.2         & 23:49:42.3     & $-$56:38:41.2     & $94\pm 41$\tablenotemark{b}	& $82\pm 8$                    & $20.3\pm 4.5$          & $5.0\pm 1.2$        \\
. . . . . . . -A               & 23:49:43.1     & $-$53:58:17.0     & $5\pm 41$\tablenotemark{b}	& $28\pm 8$	  	               & . . .                  & . . .               \\
SPT-S~J235338$-$5010.2         & 23:53:39.9     & $-$50:10:05.2     & $24\pm 21$\tablenotemark{b}	& $43\pm 6$		               & $19.9\pm 4.8$          & $5.8\pm 1.3$        \\
SPT-S~J235718$-$5153.7         & 23:57:17.1     & $-$51:53:52.1     & $74\pm 39$					& $46\pm 7$	  	               & $19.8\pm 4.6$          & $4.3\pm 1.0$        \\
\enddata
\tablecomments{SABOCA ($350\,{\rm \mu m}$) and LABOCA ($870\,{\rm \mu m}$) flux
densities of the 11 SPT sources presented in this paper, along with their
SPT ($1.4$ and $2.0{\rm \,mm}$) flux densities. The positions are the
$870\,{\rm \mu m}$ centroid positions, except for SPT\,2332$-$53-A, -B, and -C
where the $350\,{\rm \mu m}$ centroid positions are used. Unless otherwise stated, all $350\,{\rm \mu m}$ fluxes
were measured within an aperture corresponding to the LABOCA beam solid angle (see \S\ref{subsection:fluxes}). 
The errors given are statistical only. For the SED fitting analysis we add the absolute calibration uncertainty in quadrature to the statistical uncertainty.}

\tablenotetext{a}{The flux is integrated within a $25\arcsecs$ aperture covering the extended emission.} 
\tablenotetext{b}{No $350\,{\rm \mu m}$ emission peak at $S/N \ge 3$. The
$350{\rm \,\mu m}$ flux was measured within an aperture corresponding to the LABOCA beam solid angle
(see \S\ref{subsection:fluxes}).}
\tablenotetext{c}{The flux is integrated within a $44\arcsecs$ aperture covering the -A, -B, -C, and -F components.} 
\tablenotetext{d}{No $2.0\,{\rm mm}$ flux measurements available for this source, see \S\ref{subsection:fluxes} for details.} 
\tablenotetext{e}{The flux is measured within a $13\arcsecs$ aperture.} 
\end{deluxetable*}

\begin{deluxetable*}{lllcccc}
\tablecolumns{8}
\tablewidth{0pt}
\tablecaption{Redshifts and Modeled FIR Properties of SPT Sources\label{table:quantities}}
\tablehead{
Short ID$^a$                       & photo-$z$      & spec-$z^b$ &  $L_{\rm FIR}[8-1000{\rm \mu m}]$       & $SFR$                                            & $T_{\rm d}$  &   $M_{\rm d}$\\
                  &                &            &  $\times 10^{13} \Lsolar\,\mu^{-1}$     & $\times 10^{3}{\rm \Msolar\,yr^{-1}\,\mu^{-1}}$  & [K]          &   $\times 10^{10} \Msolar\,
\mu^{-1}$}
\startdata
SPT\,0512$-$59    &  $2.5\pm 1.0$  & . . .      &  $3.4\pm 1.0$                           & $5.1\pm 1.2$                                     &  ...   &   $1.0\pm 0.5$        \\
SPT\,0529$-$54    &  $2.4\pm 1.1$  & . . .      &  $4.9\pm 1.1$                           & $7.3\pm 0.7$                                     &  ...   &   $1.3\pm 0.3$        \\
SPT\,0532$-$50    &  $2.9\pm 1.2$  & . . .      &  $5.7\pm 1.2$                           & $8.5\pm 1.6$                                     &  ...   &   $1.4\pm 0.4$        \\
SPT\,0538$-$50    &  $2.5\pm 1.0$  & $2.783$    &  $6.5\pm 1.6$                           & $9.7\pm 1.5$                                     &  $34\pm 3$   &   $1.1\pm 0.2$        \\
SPT\,0550$-$53    &  $3.4\pm 1.4$  & . . .      &  $1.9\pm 0.7$                           & $2.8\pm 1.0$                                     &  ...   &   $0.6\pm 0.5$        \\
SPT\,0551$-$50    &  $2.8\pm 1.2$  & . . .      &  $2.9\pm 0.6$                           & $4.3\pm 0.6$                                     &  ...   &   $0.8\pm 0.2$        \\
SPT\,2319$-$55    &  $3.8\pm 1.6$  & . . .      &  $1.8\pm 0.5$                           & $2.7\pm 0.6$                                     &  ...   &   $0.4\pm 0.1$        \\
SPT\,2332$-$53    &  $2.5\pm 1.1$  & $2.738$    &  $6.5\pm 1.6$                           & $9.7\pm 1.3$                                     &  $36\pm 3$   &   $1.4\pm 0.3$        \\
SPT\,2349$-$56    &  $3.1\pm 1.3$  & . . .      &  $2.7\pm 0.7$                           & $4.0\pm 1.0$                                     &  ...   &   $0.9\pm 0.5$        \\
SPT\,2353$-$50    &  $4.5\pm 1.8$  & . . .      &  $1.9\pm 0.8$                           & $2.8\pm 1.2$                                     &  ...   &   $0.4\pm 0.2$        \\
SPT\,2357$-$51    &  $3.3\pm 1.4$  & . . .      &  $2.0\pm 0.7$                           & $3.0\pm 0.9$                                     &  ...   &   $0.5\pm 0.7$        \\
\enddata
\tablecomments{
Best estimates of dust temperatures, spectral indices and dust masses as
derived from the modified blackbody fits (see
\S\ref{subsection:SEDs}). The FIR luminosities, star formation rates, and dust masses are apparent and
have \emph{not} been corrected for gravitational amplification, $\mu$, and should therefore be
considered strict upper limits. For the sources with only photometric redshifts, the errors of the FIR luminosities, star formation rates, and dust masses have been derived assuming $T_{\rm d} = 34$K}
\tablenotetext{a}{Shortened source names used throughout the text, with truncated coordinates. Sources are listed in the same order as Table \ref{table:sources}}
\tablenotetext{b}{Spectroscopic redshift derived from VLT spectroscopic
observations in the restframe UV, and CO line observations with Z-Spec. For
these sources we adopt the spectroscopic redshift when calculating the $L_{\rm
FIR}$, $SFR$, $T_{\rm d}$, and $M_{\rm d}$.}
\label{table:results}
\end{deluxetable*}

\end{document}